\documentclass[12pt]{article}
\usepackage{times}
\usepackage{graphicx}
\usepackage{float}
\restylefloat{table}
\usepackage{textcomp}
\usepackage{longtable}
\usepackage{multirow}
\usepackage{enumitem}
\usepackage[top=2.0cm, bottom=2.0cm, left=2.0cm, right=2.0cm]{geometry}
\usepackage[font=small]{caption}

\makeatletter 

\newenvironment{Figure}{%
   \par\addvspace{12pt}%
   \def\@captype{figure}%
}{%
   \par\addvspace{12pt}%
}%

\long\def\@makecaption#1#2{%
  \vskip\abovecaptionskip 
  \sbox\@tempboxa{#1: #2}%
  \ifdim \wd\@tempboxa >\hsize 
    #1: #2\par 
  \else 
    \global \@minipagefalse 
    \hb@xt@\hsize{\hfil\box\@tempboxa\hfil}%
  \fi 
  \vskip\belowcaptionskip}

\makeatother 

\title{\emph{In situ} and \emph{Operando} X-ray Imaging of Directed Energy Deposition Additive Manufacturing}

\author
{Y.Chen,$^{1,2\ast}$ S. J. Clark,$^{1,2}$ L.Sinclair,$^{1,2}$ C.L.A.Leung,$^{1,2}$ S.Marussi,$^{1,2}$\\ T.Connolley,$^{3}$ O.V.Magdysyuk,$^{3}$ R.C. Atwood,$^{3}$ G.J.Baxter,$^{4}$ M.A.Jones,$^{4}$ \\ D.G.McCartney,$^{5}$ I. Todd,$^{6}$ P.D.Lee,$^{1,2\ast}$\\
\\
\small{$^{1}$Department of Mechanical Engineering, University College London,Torrington Place, London WC1E 7JE, UK.}\\
\small{$^{2}$Research Complex at Harwell, Rutherford Appleton Laboratory, Oxfordshire,OX11 0FA, UK.}\\
\small{$^{3}$Diamond Light Source, Harwell Campus, Oxfordshire, OX11 0DE, UK.}\\
\small{$^{4}$Rolls-Royce plc, PO Box 31, Derby, DE24 8BJ, UK.}\\
\small{$^{5}$Faculty of Engineering, The University of Nottingham, Nottingham, NG7 2RD, UK.}\\
\small{$^{6}$Department of Materials Science and Engineering, University of Sheffield, Sir Robert Hadfield Building,} \\ \small{ Mappin Street, Sheffield, S1 3JD, UK.}\\
\small{$^\ast$Corresponding authors: peter.lee@ucl.ac.uk, yunhui.chen@ucl.ac.uk.}
}

\date{}

\begin{document} 


\baselineskip24pt


\maketitle


\noindent \textbf{The mechanical performance of Directed Energy Deposition Additive Manufactured (DED-AM) components can be highly material dependent. Through \emph{in situ} and \emph{operando} synchrotron X-ray imaging we capture the underlying phenomena controlling build quality of stainless steel (SS316) and titanium alloy (Ti6242 or Ti-6Al-2Sn-4Zr-2Mo). We reveal three mechanisms influencing the build efficiency of titanium alloys compared to stainless steel: blown powder sintering; reduced melt-pool wetting due to the sinter; and pore pushing in the melt-pool. The former two directly increase lack of fusion porosity, while the later causes end of track porosity. Each phenomenon influences the melt-pool characteristics, wetting of the substrate and hence build efficacy and undesirable microstructural feature formation. We demonstrate that porosity is related to powder characteristics, pool flow, and solidification front morphology. Our results clarify DED-AM process dynamics, illustrating why each alloy builds differently, facilitating the wider application of additive manufacturing to new materials.}

\clearpage

\section*{Introduction}

Laser Additive Manufacturing (LAM)({\it 1}) is transforming modern manufacturing technologies and enabling direct fabrication of complex shapes({\it 2}) for metals({\it 3}), ceramics({\it 4}) and polymers({\it 5}) using computer-aided designs. This level of flexibility and customizability({\it 6}) cannot be achieved through traditional manufacturing routes. Directed Energy Deposition Additive Manufacturing (DED-AM) is amongst the most promising methods in LAM due to its capabilities from repair applications to direct build of functionally-graded and large freeform components. In contrast to Laser Powder Bed Fusion (LPBF) systems, the DED-AM process (also known as the blown powder process) involves the feeding of powder materials from a nozzle into a laser-heated molten metal pool. However, the lack of fundamental understanding of the underlying process–structure-property relationships hinders the utilisation of DED-AM for production. Technical challenges include the accumulation of significant residual stresses({\it 7}) and formation of undesirable microstructures({\it 8}), which can restrict the mechanical performance of the final product. Detrimental features such as gas porosity, lack of fusion porosity and balling of the melt-track can occur and these can have a deleterious effect on the structural integrity of the final components. Specifically, DED-AM is an immensely attractive route for the manufacture of titanium alloy aerospace parts({\it 9}) because customised near-net-shape components can be produced without the need for machining which is both time consuming and wasteful of an expensive metallic alloy. However, it has been widely reported that it is difficult to achieve the structural integrity required and this has proved to be a major barrier to more widespread application. \\

\noindent The majority of the experimental studies of the DED-AM processes have involved \emph{ex situ} characterisation of the microstructure({\it 10}) and the mechanical properties({\it 11}). However, this approach does not provide a fundamental understanding of the mechanisms by which microstructural features form. \emph{In situ} studies, utilizing infrared imaging, high-speed optical imaging, thermocouples and pyrometers, to develop closed-loop control({\it 12}) of the manufacturing process have been reported but these also fail to provide fundamental insights. On the other hand, \emph{in situ} and \emph{operando} high-speed X-ray investigations({\it 13}) using high-flux third-generation synchrotron radiation sources({\it 14}) have enabled researchers to see into the heart of the powder bed AM process({\it 15}) and characterise the transient phenomena({\it 16}) in the melt-pool({\it 17}) such as flow({\it 18}), microstructural feature formation({\it 19}) together with phase and strain evolution({\it 20}). This capability is due to the ability of X-rays to penetrate through a sufficient depth of dense, optically opaque metallic samples with high spatial and temporal resolution({\it 21}). Therefore, such an approach has the potential to help elucidate the phenomena occurring on the time and length scales associated with DED-AM. Although synchrotron radiography has been employed to study laser powder bed fusion (LPBF), much less attention has been given to DED-AM process. Wolff et al.({\it 19}) simulated some aspects of the blown powder process by using a piezo-driven vibration-assisted powder delivery system to induce a gravitational flow of powders from a syringe-needle; however, their process replicator was vastly different from an industrial DED-AM process due to the low-velocity unfocused powder delivery and very finely focused laser. \\

\noindent In this work, we perform a high-resolution, time-resolved \emph{in situ} and \emph{operando} synchrotron X-ray radiographic study of DED-AM using an AM machine with build capability that directly scales to common industrial DED-AM processes. Here we report the investigation of deposition phenomena in different materials including melt-pool dynamics and the relationship to microstructure formation mechanisms in a multi-layer blown powder build. We successfully capture the laser-matter interactions during DED-AM for SS316 (grade 316 iron-based austenitic stainless steel) and the aerospace titanium alloy Ti6242 (Ti-6Al-2Sn-4Zr-2Mo). These observations reveal the vastly different physical phenomena occurring during DED-AM for different materials, including the underlying physics causing melt-track formation mechanisms to be different in the two alloys. Compared with SS316, we uncovered that during the processing of Ti6242, blown powder sinters ahead of, and behind, the melt-pool. The presence of this has a significant influence on the development of the build and on the mechanisms of porosity formation. The results presented in this work provide an enhanced fundamental understanding of the DED-AM process with direct relevance to optimising the microstructures of titanium alloy components.

\section*{Results \& Discussion}

\subsection*{1. \emph{In situ} and \emph{operando} synchrotron X-ray imaging of DED-AM}

We performed \emph{in situ} and \emph{operando} X-ray imaging on the $I12$ Joint Engineering, Environmental, and Processing (JEEP) beamline({\it 22}) at Diamond Light Source to capture the transient phenomena in the DED-AM of SS316 and Ti6242 (see Supplementary Figure 1). The Blown Powder Additive Manufacturing Process Replicator (BAMPR) was designed to replicate a scaled version of commercial DED-AM systems, enabling integration in synchrotron beamlines, as shown in Figure. 1a (and Suppl. Figure 1). Further details about this bespoke instrument can be found in methods and Supplementary Information. Specifically, the instrument facilitates deposition with a laser power density of up to 6.37 × 10$^{3}$ $W\ mm$$^{-2}$ and a corresponding energy density of up to 32 $J\ mm$$^{-3}$. These specifications were designed to faithfully replicate industrial conditions (see Supplementary Table 1-3 for a detailed comparison to the reported parameters for commercial systems). \\

\noindent A schematic of the X-ray imaging process is shown in Figure. 1b. The synchrotron X-rays are attenuated by the powder and deposited material, which is then converted to visible light via a scintillator and captured as a radiographic video using a fast CCD camera (not shown). As opposed to LPBF({\it 23}), the weld pool in DED-AM is significantly larger ($mm$ compared to 200 $mm$) and the traverse speed slower ($mm\ s$$^{-1}$ as compared to $m\ s$$^{-1}$). Furthermore, the melt-pool is stationary with respect to the camera’s frame of reference. Therefore, the optimal acquisition speed for imaging the weld pool dynamics was selected as 200 $fps$ (with 0.0049 $s$ exposure time). This framerate provided sufficient signal to noise ratio to resolve the subtle solid-liquid phase contrast boundary. For the experiments capturing the fast-transient influence of powder on melt-pool initiation 2000 $fps$ was used with a subsequent trade-off in signal quality. \\

\begin{Figure}
\includegraphics[width=14cm]{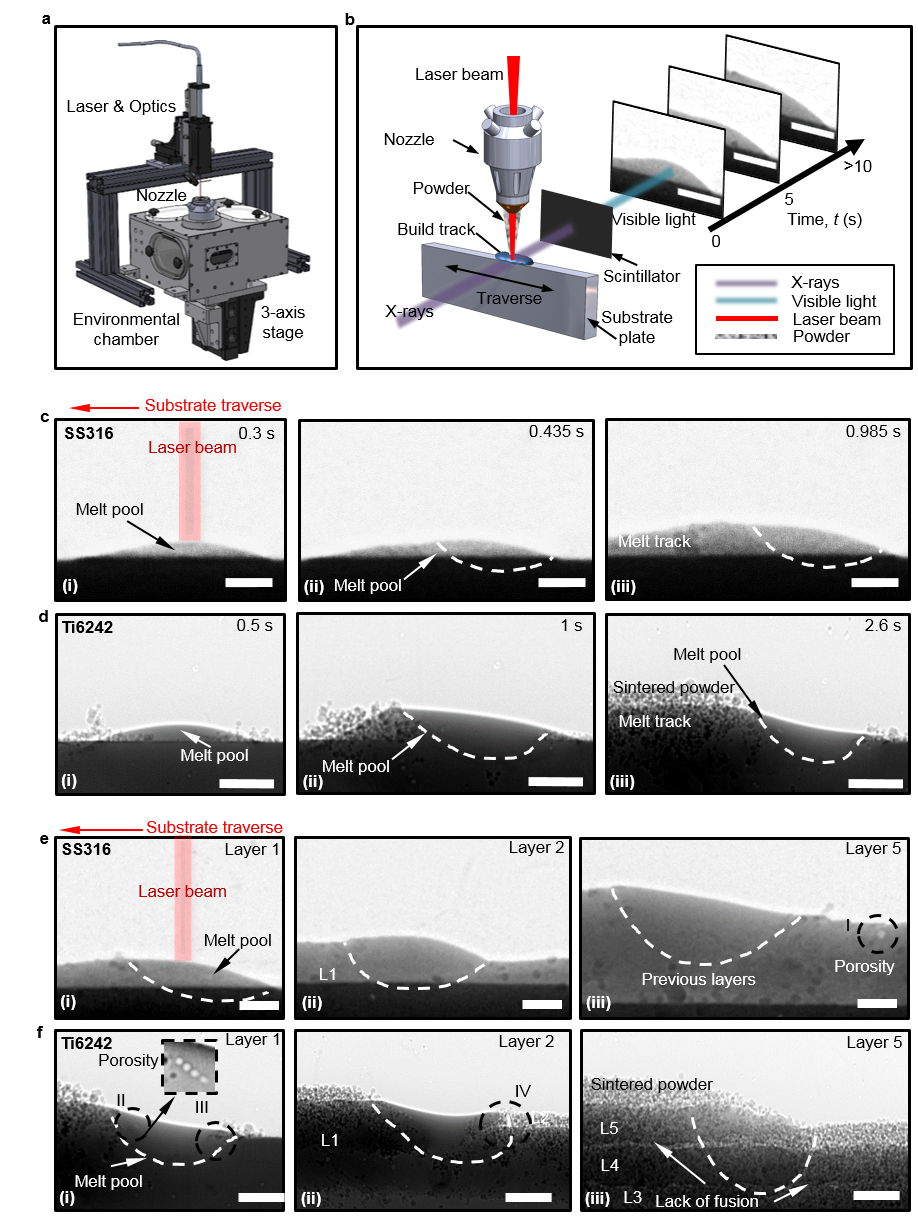}
\centering
\caption{ \textbf{\emph{In situ} and \emph{operando} X-ray imaging of the DED-AM process.} (A) The Blown Powder Additive Manufacturing Process Replicator (BAMPR) replicates a commercial DED-AM system. The system is encapsulated within a Class $I$ laser enclosure and comprises an inert environmental chamber, a high precision 3-axis platform, an industrial coaxial DED nozzle and a laser system. (B) Schematic of the \emph{in situ} and \emph{operando} X-ray imaging of DED-AM process. The synchrotron X-rays are attenuated by the deposited materials and then are converted to visible light with a scintillator to capture as a radiographic video on a fast CCD camera (not shown). The resulting videos reveal the underlying mechanism and dynamics of the process. Scale bars on the radiographs = 1000 $\mu m$. (C) Time-resolved radiographs acquired during DED-AM at the start of a melt-track on SS316 (Supplementary Video 1) and Ti6242 (Supplementary Video 2). SS316 build is under laser power density ($P$) = 6.37 × 10$^{3}$ $W\ mm^{-2}$, scanning speed ($v$) = 1.67 $mm\ s^{-1}$, laser spot size = 200 $\mu m$. Ti6242 build is under P = 1.59 × 10$^{3}$ $W\ mm^{-2}$, $v$ = 1 $mm\ s^{-1}$, laser spot size = 400 $\mu m$. Both captured at 200 $fps$, Scale bar = 500 $\mu m$. Three radiographs were selected as representative of the building process. The time since the build started is marked on each radiograph. Agglomerated/sintered particles can be observed on top of the Ti6242 melt-track during the building process. (D) Time-series radiographs acquired during DED of an SS316 (Supplementary Video 3) and a Ti6242 (Supplementary Video 4) multi-layer build. Three radiographs were selected as representative of the building process of layer 1, layer 2 and layer 5, respectively. A pore (broken circle) is observed in layer 5 of the SS316 build and numerous pores are observed in the Ti6242 build. The sintered layer caused lack of fusion between layers of Ti6242. Scale bar = 500 $\mu m$. The melt-pool boundary in (C) and (D) are marked by monitoring the solid-liquid moving boundary.}
\end{Figure}

\subsection*{2. Melt-track Initiation}

To reveal the initial establishment of a melt-track at the onset of the build was captured at varying laser powers of 50 $W$, 100 $W$ and 200 $W$ (laser power density 1.59 × 10$^{3}$ $W\ mm$$^{-2}$, 3.18 × 10$^{3}$ $W\ mm$$^{-2}$ and 6.37 × 10$^{3}$ $W\ mm$$^{-2}$, respectively with a laser spot size of 200 $\mu m$) (Figure 2a, b \& c and Supplementary Videos 5, 6 \& 7 respectively) at an image acquisition rate of 2000 fps. We applied a local-temporal background subtraction to emphasise the moving objects in the radiographs detailed in the supplementary information. In Figure 2a, with a laser power of 50 $W$, a melt-track was not observed to form, little thermal expansion of the substrate was evident, and powder particles were not seen to adhere to the substrate surface. In contrast, as shown in Figure 2b, upon increasing the laser power to 100 $W$ the initiation of the particle incorporation process becomes evident. Powder particles are observed to adhere onto the substrate plate commencing at 1.065 s after the laser was turned on, suggesting the surface is molten. These powder particles coalesce and form a melt-pool after 1.125 $s$. With a further increase of laser power to 200 $W$, as shown in Figure 2c, the radiographs show that just 0.0695 $s$ after the laser was turned on, a powder particle was captured and incorporated into the melted substrate area and formed a melt-pool after just 0.131 $s$. It is evident at both powers that the adhesion of the powder particles significantly increases the absorption of the laser energy. We hypothesise this is due to the increased surface area. \\

\begin{Figure}
\includegraphics[width=12cm]{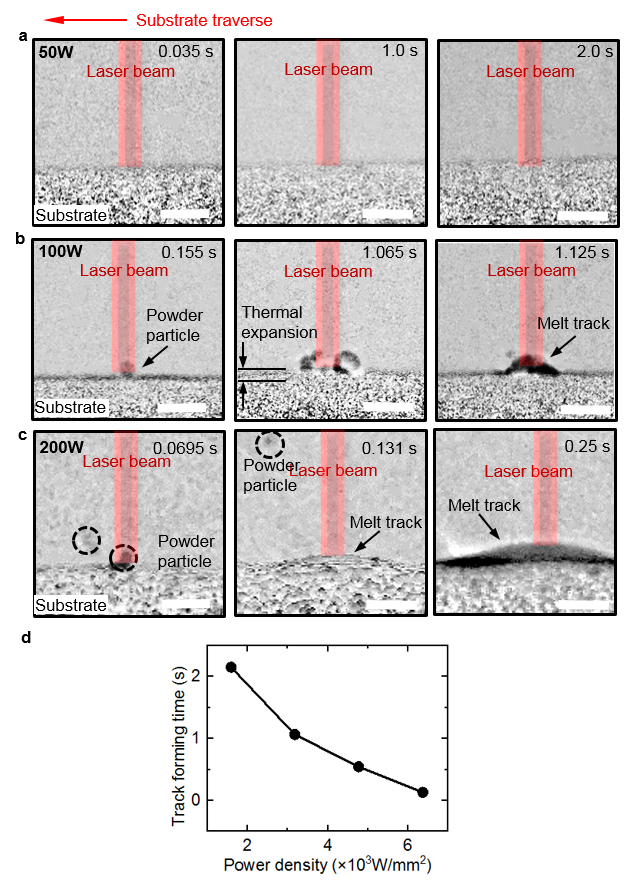}
\centering
\caption{ \textbf{Time-series radiographs acquired during DED-AM revealing the mechanism of melt-track formation. } Material is SS316. All radiographs were captured at 2000 $fps$ and revealed by local-temporal background subtraction. (A) P = 6.37 × 10$^{3}$ $W\ mm$$^{-2}$, v = 1.67 $mm\ s$$^{-1}$. See Supplementary Video 5. Scale bar = 1000 $\mu m$. (B) P = 3.18 × 10$^{3}$ $W\ mm$$^{-2}$. See Supplementary Video 6. Scale bar = 500 $\mu m$. The powder particles agglomerated/sintered under laser radiation and eventually forming a melt-track via surface wetting. Powder particles were wetted onto the melted substrate plate separately and deposited onto the substrate surface at 1.065 $s$ after the laser was turned on. Three powder particles coalesced at 1.065 $s$ and formed a melt-track at 1.125 $s$ with more powder particles coalesced with the melt-track. (C) P = 1.59 × 10$^{3}$ $W\ mm$$^{-2}$. See Supplementary Video 7. Scale bar = 500 $\mu m$. (D) The relationship between the time a melt-track needs to form after the laser turns on and laser power density. }
\end{Figure}

\noindent The DED-AM process exhibits a complex laser-matter interaction dependent upon multiple factors such as; surface roughness, angle of incidence and the presence of surface oxides({\it 24}). The melt-track initiation mechanism is a transient process which occurs when the absorbed power density exceeds a critical power density threshold. While the melt-track development mechanism here is based on laser power density which operates in the melting region({\it 25}). The relationship between the time a melt-track needs to form after the laser turns on and laser power density is plotted in Figure 2d. When the laser illuminates the substrate surface with a power density of 6.37×10$^{3}$ $W\ mm$$^{-2}$, instead of forming a ‘cavity’(19) or ‘keyhole’({\it 26}), as can often occur in the LPBF({\it 27}) process, the substrate surface is melted to give a ‘conduction mode’({\it 28}) melt-pool. In the present set-up, the powder particle flow is concentric with the laser beam and therefore the powder particles are directly deposited into the region the melted region. Some of these particles can be incorporated into the melt-pool, driven by surface tension and capillary forces, however, others rebounded where they impact solid material. Decreasing the laser power density slows the met-track initiation process as the molten area of the substrate is both narrower and shallower reducing both the probability of a powder particle impacts the molten area and that the powder particle is in-turn incorporated into the pool. The reduction of laser powder density therefore, directly results in a corresponding reduction of powder capture efficiency.\\

\subsection*{3. Powder sinter}

The time-resolved evolution of the multi-layer melt-track morphology of an SS316 track on an SS304 substrate and a Ti6242 track on a titanium Ti-6Al-2Sn-4Zr-6Mo (Ti6246) substrate, respectively, were captured by X-ray at 200 $fps$ during the DED-AM laser deposition process (Figure 3c \& d and Supplementary Videos 1-4). Single melt-tracks were deposited in an alternating bi-directional strategy up to 10 layers in height, with the melt-pool measuring 1 to 3 $mm$ in diameter (see Supplementary info). Figure 1c shows a time-series of radiographs taken from the start of the deposition process. The laser beam is shown to melt the substrate surface and consolidate the powder particles into a melt-pool and form a continuous deposited track during the deposition of the first layer. The extent of the melt-pool is marked with a dotted line in Figure 3c-d, and can be observed clearly in Supplementary Videos 1-4 since the liquid is less dense and hence brighter than the solid in radiography; there is also a phase-contrast fringe at the solid-liquid boundary. It should be noted that the shape of the melt-pool is reversed from that of a classic convex pool penetrating the track below (as in the SS316 case), to pool resembling a saddle (see Supplementary Figure 5). The saddle shape leads to the appearance of 2-3 phase contrast fringes generated by the local saddle-points through projection in the X-ray direction. \\

\noindent An additional significant difference in the melt-track morphology between SS316 and Ti6242 can be observed from the radiographs, as shown in Figure 1c and 1d, in that sintered powder particles are observed covering the melt-track and around the melt-pool during the multi-layer build of Ti6242. The thickness of the sintered powder can reach several times that of the melt-track itself as shown in more detail in Figure 3e and Supplementary Video 8. The extent of the sintering appears to depend strongly on the input laser power density. In subsequent layers of the build, the slow percolation of the melt-pool through the sintered powder layer further exacerbates the inversion of melt-pool, causing it to form a saddle shape. This phenomenon is captured in Figure 3b where the front of the melt-pool exhibits a ‘crocodile’s mouth’ shape during the second layer of the build. This key difference between SS316 and Ti6242, i.e. the formation of sinter, is illustrated schematically in Figure 3c and d where infiltration of the Ti6242 melt into the porous powder layer is emphasised and compared. \\

\begin{Figure}
\includegraphics[width=12cm]{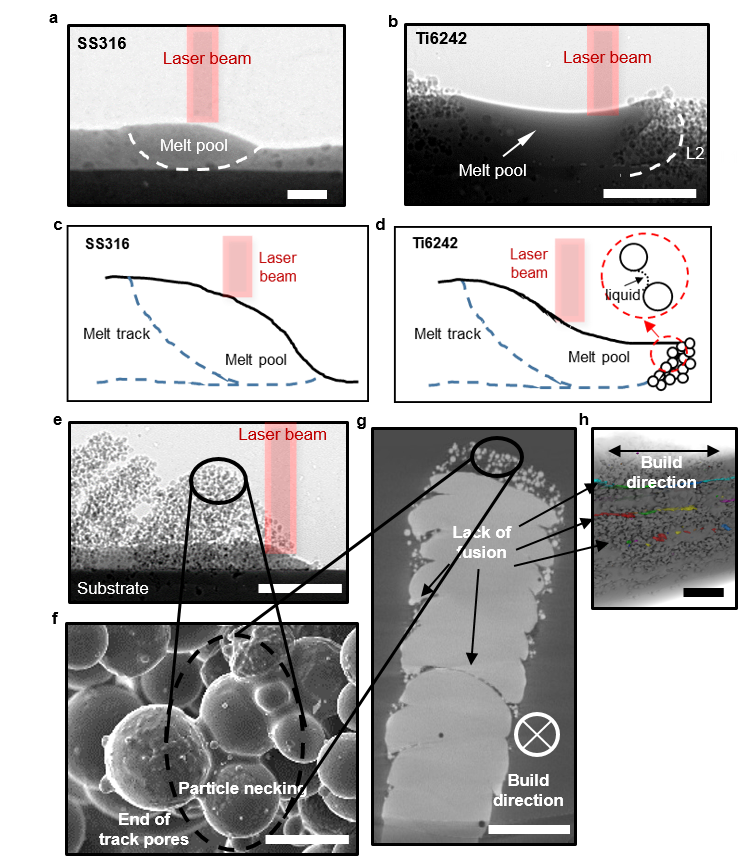}
\centering
\caption{ \textbf{Understanding the powder sintering phenomenon in Ti6242. } (A) Time-resolved radiograph acquired during an SS316 build revealing the convex shape of the melt-pool surface, Laser power density = 6.37 × 10$^{3}$ $W\ mm$$^{-2}$, v = 1.67 $mm\ s$$^{-1}$ captured at 200 $fps$. Scale bar = 1000 $\mu m$.  in comparison to (B) Time-resolved radiograph acquired during a Ti6242 build revealing the concave shape of the melt-pool surface due to the sintered powder layer, Laser power density = 1.59 × 10$^{3}$ $W\ mm$$^{-2}$, v = 1 $mm\ s$$^{-1}$ captured at 200 $fps$. Scale bar = 1000 $\mu m$. (C) Schematic of (a) showing the convex shape melt-pool in a multi-layer build condition. (D) Schematic of (B) showing the phenomena in a multi-layer build condition, the percolation of the melt-pool through the loosely sintered powder layer distinctively alters the melt-pool geometry.  Laser power density = 1.59 × 10$^{3}$ $W\ mm$$^{-2}$, v = 1 $mm\ s$$^{-1}$ captured at 200 $fps$. Scale bar = 1000 $\mu m$. (E) Time-resolved radiograph showing significant powder agglomeration/sintering during Ti6242 build, Laser power density = 7.96 × 10$^{2}$ $W\ mm$$^{-2}$, v = 1 $mm\ s$$^{-1}$ captured at 200 fps. See Supplementary Video 8. The sintered layer thickness is more than three times the height of the melt-track itself. Scale bar = 1000 $\mu m$. (F) Post-mortem SEM image of Ti6242 partially sintered particle necks attached to the melt-track of (E). Diffusion provides the mechanism for the flow of material from regions of low curvature to regions of high curvature i.e. towards the neck or point of contact. Scale bar = 10 $\mu m$. (G) Cross-section of the melt-track from the micro-CT scan of Ti6242 melt-track revealing the lack of fusion between layers and the sinter particles surrounded. Scale bar = 100 $\mu m$. (H) Post-mortem micro-CT scan of Ti6242 melt-track revealing lack of fusion porosity between layers caused by sintering phenomenon. Scale bar = 500 $\mu m$. }
\end{Figure}

\noindent To understand more about the sintered Ti6242 powder, the partially sintered layer attached to the melt-track was examined in a scanning electron microscope and inter-particle necks are revealed, as shown in Figure 3f and Supplementary Figure 7. The capture efficiency by the melt-pool of powder particles heated in the laser beam is less than one({\it 29}). Particles that are not captured by the melt-pool can potentially stick to the surface of the substrate$/$solidified track; a physical phenomenon that is known to occur in other processes for the particle sizes ($\sim$50 $\mu m$) and the velocities ($\sim$5 $m\ s^{-1}$) similar to those in DED ({\it 30}). However, with Ti6242 it appears to actually sinter. One hypothesis for the formation of Ti6242 powder sinter is that their surface oxide dissolves into the alloy as the powder surface heats due to the increased solid solubility of oxygen at temperatures over 1000 $^\circ C$ ({\it 31}). The exposed high temperature alloy bonds quickly to other particles as seen during Ti alloy sinter treatments ({\it 32}), while for SS316, the powder has a thin chromia-type oxide that is strongly adherent and chemically stable({\it 33}) even when heated. I.e. the SS316 oxide acts as a barrier to sticking and interdiffusion between particles. Hence impinging Ti6242 particles are very likely to stick to one another and, without an oxide diffusion barrier, interdiffusion will occur across clean metal interfaces causing a particle neck to form, as shown in Figure 3d. \\

\noindent The lightly sintered, very porous layer can potentially causes lack of fusion porosity between layers in a multi-layer DED build, Figure 3d, Figure 3g and h. Detailed examples of the formation of porosity and lack of fusion features can be found in Supplementary Figure 8. \emph{In situ} radiography shows that the sinter layer can alter melt-pool, affect both surface roughness and lack of fusion porosity in the titanium alloy builds.\\

\subsection*{4. Porosity formation}

Porosity can occur in DED-AM builds and can be categorised into lack of fusion and gas entrapped pores({\it 34}). We have already discussed lack of fusion pores. With regards to gas entrapped pores, previous research has suggested they are  generated via two sources: laser-induced keyhole porosity and gas entrapped porosity due to absorption of gas from the feedstock powder({\it 19}). It is reported that a keyhole can trap shielding gas during DED-AM and contribute to gas entrapped pore formation({\it 19,26}). However, DED-AM is not normally associated with ‘keyhole mode’ and this phenomenon was not observed in our experiments. Here, we report that entrained porosity (Figure 4a and 4b and Supplementary video 9 \& 10) is related to gas in the feed powder (entrained and soluble) and the influence of Marangoni flow.\\

\noindent A times series of radiographs (Figure 4a) suggest a potential mechanism of sole pore formed during a 10 layer SS316 build. At t = 0.01 $s$ into the build, a cluster of 3-4 agglomerated powder particles hits the leading edge of the pool and melts into it. A pore then appears to form at the same location (t = 0.015 $s$), perhaps entrained from this powder cluster. This gas pore then swirls around several times in the melt-pool, probably following the Marangoni flow streamlines. After a third of a second, the pore reaches the rear of the pool, and is entrained in the semi-solid, forming a pore in the solid track (t = 0.315 $s$, Figure 4a). \\

\noindent In contrast, in Ti6242 many pores are formed, as is typical for Ti alloy welds. We hypothesise that there was trapped gas within the sintered powder layer which originated from either the powder particles themselves({\it 35}) or from the porous sinter layer. Interestingly, unlike the pore in SS316, when the pores in Ti6242 reach the back, they are pushed along rather than captured by the solid. The actual flow is quite complex, as the pores are in the rear of the two pools on either side of the saddle shape (see Supplementary Figure 5), one larger and deeper than the other. These two clusters of pores hover in the pool below the surface, as the downward Marangoni flow pushes them down balanced against the buoyancy force. We hypothesise that these pores are pushed by the solidification front rather than entrained due to Ti6242’s very narrow freezing range ($\sim$5 $K$)(36), which favours a planar solidification front ({\it 36}) at the traverse speed of 1 $mm\ s^{-1}$ $-$ 10 $mm\ s^{-1}$ (see Figure 4c and Supplementary Figure S6). The planar solidification front (which cannot be resolved in the present experiments and is not preserved to room temperature due to the solid-state phase transformation({\it 37})) effectively prevents the pores from being engulfed into the solidifying track and instead results in pore pushing. The pores are thus swept along with the melt-pool at its back wall (Figure 4b) until the end of the track is reached, as shown in the XCT scans in the Supplementary Figure S7.\\

\begin{Figure}
\includegraphics[width=13cm]{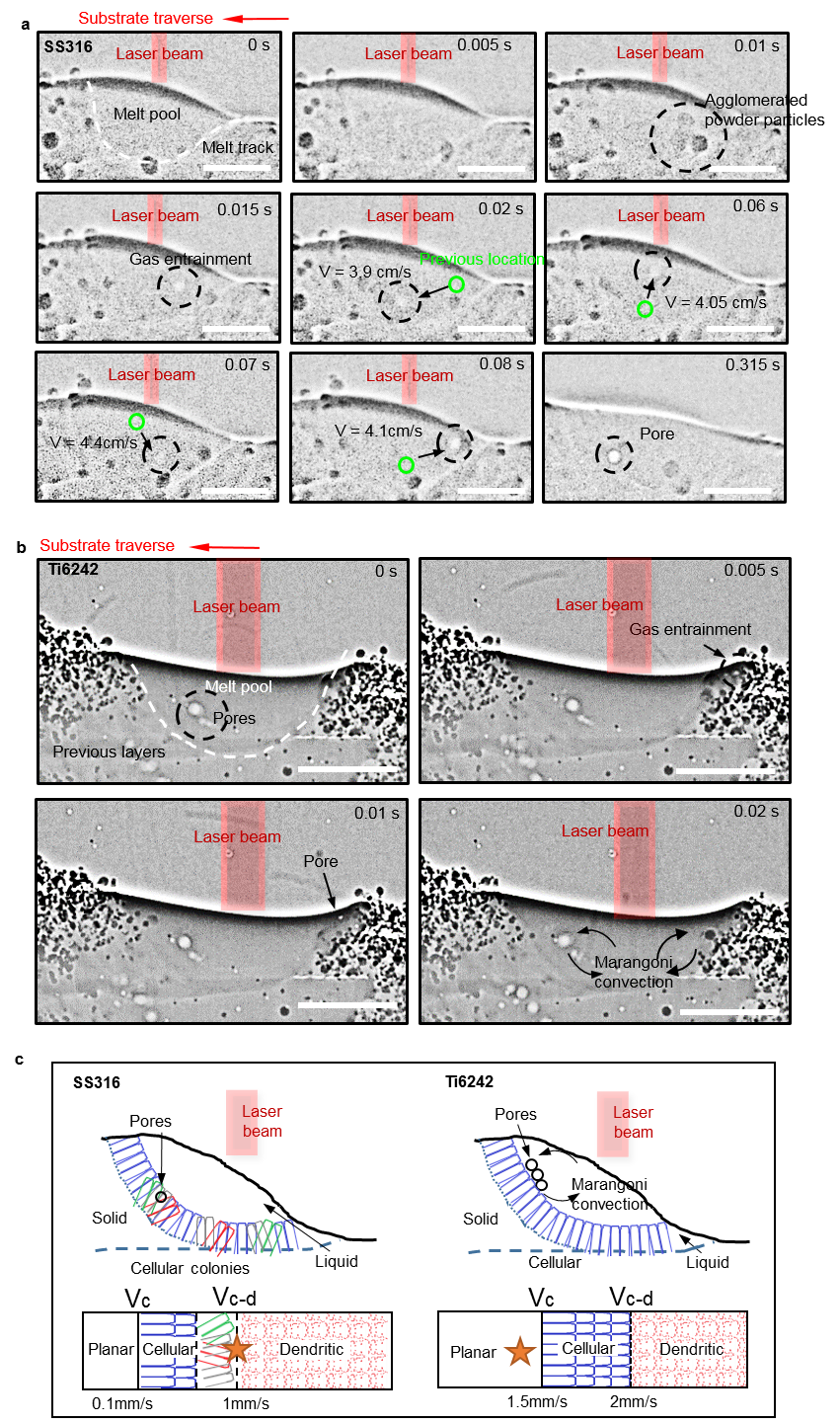}
\centering
\caption{ \textbf{Porosity formation and pore pushing mechanisms captured during X-ray imaging. } (A) SS316: A pore appears to form via gas entrainment from a cluster of powder particles melted into the pool. A pore flow trajectory revealed on a radiograph. See Supplementary Video 9. The Marangoni-driven flow forces the pore to flow in the melt-pool at an average velocity of $\sim$ 43 $mm\ s^{-1}$. Scale bar = 1000 $\mu m$. (B) Ti6242: a pore appears to form via gas entrainment from the sintered particles ahead of the melt-pool. See Supplementary Video 10. Pores are being swept along with the melt-pool following the convection flow. Scale bar = 1000  $\mu m$. (C) Schematic of the pore formation mechanisms with a comparison between Ti6242 and SS316. The planar/cellular front in Ti6242 effectively prevents the pores from being engulfed into the solidifying track. As a result, the pores at the solid/liquid interface at the back of the melt-pool are swept along with it. In SS316, a cellular colony solidification structure develops which engulfs the pores during solidification. This results in the formation of a series of pores that are regularly distributed along the length of the melt-track. }
\end{Figure}

\noindent In contrast to Ti6242, SS316 has a much wider freezing range ($\sim$110 $K$)({\it 38}) and therefore has a cellular colonies solidification morphology. This morphology tends to allow pores to be engulfed by the solidification front and not pushed towards the end of the track, as shown in Figure 4c and Supplementary Figure S6. In SS316, this gives rise to the formation of pores that are entrained shortly after they enter the melt pool, and hence are more regularly distributed along the length of the melt-track, unlike Ti6242 where end of track agglomeration occurs (Supplementary Figure S8). \\

\section*{Summary}

We have used fast time-resolved synchrotron X-ray imaging to uncover fundamental physical phenomena during the multi-layer DED-AM process \emph{in situ} and \emph{operando} in that dramatically alter the behaviour of different key engineering alloys. We reveal insights into the underlying mechanisms governing melt-track formation, melt-pool dynamics and transient porosity formation. We show these mechanisms are highly material dependent, differing significantly between SS316 and Ti6242. Using different laser power densities, we demonstrate that the mechanism for initiation and formation of a melt-track involves laser-induced surface wetting driven by capillary forces. We observed a significant difference between SS316 and Ti6242; the latter forms an agglomerated$/$sintered powder layer on the surface of a track. We conclude that the formation of a sintered layer in Ti6242, which is not found in SS316, is due to differences in the behaviour of their surface oxides. The alteration of the melt-pool spreading due to the presence of the loosely sintered layer is the main cause of high surface roughness and lack of fusion pores in the final product of Ti6242. \\

\noindent We capture the different pore formation mechanisms in SS316 and Ti6242 builds and conclude that porosity is by gas entrainment from the powder feedstock. In Ti6242 pores are observed to swirl in the melt-pool and are not entrained into the solidification front most probably due to it being planar. Instead, the pores are pushed, forming a cluster of pores at the end of the track. In contrast, SS316 has a much wider freezing range giving a dendritic solidification morphology which readily entrains the pores rather than pushing them, resulting in a more even distribution of pores along the length of the track. \\

\noindent Overall, our results clarify the key fundamental physics behind DED-AM process and can be used to design processing strategies for enhanced quality and performance of additively manufactured parts in stainless steel, titanium and other alloys.\\

\section*{Materials and Methods}
\subsection*{1. \emph{In situ} and \emph{operando} synchrotron X-ray imaging}

The BAMPR (Blown Powder Additive Manufacturing Process Replicator) has been developed to faithfully replicate a commercial DED-AM system which used industrial power density and energy density standards. The instrument was specifically designed to be integrated into high energy high flux synchrotron beamlines, $e.g.$ $I12$: Joint Engineering, Environmental, and Processing (JEEP) at Diamond Light Source. $I12$ is chosen for its high energy range (53-150 $keV$) and well-suited for the size of the melt-pool and the thickness of the melt-track (1-3 $mm$ thick). Due to the industrial scale of the experiment and the limitation on the flux, it is proven difficult to increase to frame-rate to more than 2,000 $fps$. The system is encapsulated within a Class $I$ laser enclosure and comprises an inert environmental chamber, a high precision 3-axis stage, an industrial coaxial DED nozzle and a laser system. An industrial powder feeder (Oerlikon Metco TWIN-10-C) delivers powder to the system in a stream of argon gas. The powders used were gas atomised Ti6242 and SS316 powders with a size distribution of 45 $\mu m$ - 90 $\mu m$ and with median particle diameter, $D_{50}$, of 70 $\mu m$. The laser is a 1070 $nm$ Ytterbium-doped fibre laser beam (continuous-wave ($CW$) mode, laser power ($P$) of 0 $W$ - 200 $W$) which is coupled with tuneable optics to facilitate a focused spot size of between 200-700 $\mu m$. The laser is positioned to be concentric with the powder delivery stream blown from the nozzle and normal to the substrate plate (1-2.5 $mm$ thick). As a result, the melt-pool is steadily placed in the centre of the X-ray beam during material deposition process. This provides huge advantages in observing melt-pool dynamics. Ultra-high time resolution is not necessary for our experiments due to its disadvantage in recording powder flow and over-shadow the melt-pool.  In this work, the 3-axis stage was able to translate through 25 $mm$ in length, 50 $mm$ in width, and 50 $mm$ in height. The powder used was gas atomised Ti6242 and SS316 powders with a size distribution of 45 $\mu m$ - 90 $\mu m$ with median particle diameter, $D_{50}$, of 70 $\mu m$. The substrate plate is positioned inside the environmental build chamber with Kapton film (X-ray transparent) windows under a flowing argon atmosphere of 6 $L\ min^{-1}$. The speed of the sample stages in both cases was controlled to be 1-5 $mm\ s^{-1}$ to enable a continuous track to be formed for this condition as indicated by preliminary laboratory trials.\\

\subsection*{2. Image processing and quantification}
The acquired images are corrected to form a Flat-Field Corrected (FFC) image by dividing by an average of 100 flat field images using $ImageJ$ and $Matlab$\textcircled{c}. To increase the image contrast and signal-to-noise ratio, we applied a local-temporal background subtraction to reveal key information in the X-ray images following equation:
\begin{equation}
\label{Eq1}
	LTBS = \frac{FFC}{I_{lavg}}
\end{equation}
Where $LTBS$ is the local-temporal background-subtracted image, $FFC$ is the flat field corrected image, and $I_{lavg}$ is a local average of 50 of the nearest neighbour images (25 before and 25 after). The “local-temporal background subtraction” method is chosen to reveal the subtle solid-liquid interface at the boundary of the melt-pool in sacrifice of time resolution. In the high image resolution experiment, 200 $fps$ was used to maximise the spatial resolution in the experimental set-up and to avoid the smudge of images from the fast-moving powder particles.

\section*{Acknowledgments}
\noindent \textbf{General:} The authors also would like to thank Dr. Ali Dehghan-manshadi for performing the microstructure characterization on SS316 samples.\\
\noindent \textbf{Funding:} This research was supported under MAPP: EPSRC Future Manufacturing Hub in Manufacture using Advanced Powder Processes (EP/P006566/1), a Royal Academy of Engineering Chair in Emerging Technology, and Rolls-Royce plc. via the Horizon 2020 Clean Sky 2 WP5.8.1 programs and through support of L.S.’s studentship. S.C. was supported by the Office for Naval Research under grant number N62909-19-1-2109.\\
\noindent \textbf{Author contributions:} P.D.L. conceived the project. Y.C., S.C. and S.M., led the design of the Blown Powder Additive Manufacturing Process Replicator (BAMPR). Y.C. and S.C. designed and performed the experiments, with all authors contributing. Y.C. performed the data analysis with S.C contributing. Y.C and P.D.L. led the results interpretation and paper writing, with all authors contributing.
     
\section*{References}
\begin{itemize}
\item[1.]
T. Debroy, H.L. Wei, J.S. Zuback, T. Mukherjee, J.W. Elmer, J.O. Milewski, A.M. Beese, A. Wilson-Heid, A. De, W. Zhang, {\it Prog. Mater. Sci.} {\bf 92}, 112-224 (2018). 
\item[2.]
J. Beuth, J. Fox, J. Gockel, C. Montgomery, R. Yang, H. Qiao, E. Soylemez, P. Reeseewatt, A. Anvari, S. Narra, N. Klingbeil, {\it 24th Int. SFF Symp. - An Addit. Manuf. Conf.} {\bf SFF 2013}, 655-665 (2013). 
\item[3.]
T. B. Sercombe, G. B. Schaffer, {\it Science} {\bf 301}, 1225–1227 (2003). 
\item[4.]
Z. C. Eckel, C. Zhou, J. H. Martin, A. J. Jacobsen, W. B. Carter, T. A. Schaedler, {\it Science} {\bf 351}, 58-62 (2016). 
\item[5.]
M. Schmid, K. Wegener, {\it Procedia Eng.} {\bf 149}, 457–464 (2016). 
\item[6.]
E. MacDonald, R. Wicker, {\it Science} {\bf 353}, 655-665 (2016). 
\item[7.]
M. Megahed, H.-W. Mindt, N. N’Dri, H. Duan, O. Desmaison, {\it Integr. Mater. Manuf. Innov.} {\bf 5}, 61-93(2016). 
\item[8.]
N. Shamsaei, A. Yadollahi, L. Bian, S. M. Thompson,{\it Addit. Manuf.} {\bf 8}, 12-35 (2015). 
\item[9.]
T. M. Pollock,  {\it Nat. Mater.} {\bf 15}, 809-815 (2016). 
\item[10.]
C. Kumara, A. Segerstark, F. Hanning, N. Dixit, S. Joshi, J. Moverare, P. Nylén, {\it Addit. Manuf. } {\bf 25}, 357-364 (2019). 
\item[11.]
A. Alafaghani, A. Qattawi, M. A. G. Castañón, {\it Int. J. Adv. Manuf. Technol.} {\bf 99}, 2491-2507 (2018). 
\item[12.]
Y. Miao, T. Yao, J. Lian, J. S. Park, J. Almer, S. Bhattacharya, A. M. Yacout, K. Mo,  {\it Scr. Mater.} {\bf 131}, 29-32 (2017). 
\item[13.]
A. A. Martin, N. P. Calta, S. A. Khairallah, J. Wang, P. J. Depond, A. Y. Fong, V. Thampy, G. M. Guss, A. M. Kiss, K. H. Stone, C. J. Tassone, J. Nelson Weker, M. F. Toney, T. van Buuren, M. J. Matthews, {\it Nat. Commun. } {\bf 10}, 1-10 (2019). 
\item[14.]
A. S. Schlachter, {\it New Directions in Research with Third-Generation Soft X-Ray Synchrotron Radiation Sources}, 1-22 (1994). 
\item[15.]
N. D. Parab, C. Zhao, R. Cunningham, L. I. Escano, K. Fezzaa, W. Everhart, A. D. Rollett, L. Chen, T. Sun, {\it J. Synchrotron Radiat.} {\bf 25}, 1467-1477 (2018). 
\item[16.]
S. A. Khairallah, A. A. Martin, J. R. I. Lee, G. Guss, N. P. Calta, J. A. Hammons, M. H. Nielsen, K. Chaput, E. Schwalbach, M. N. Shah, M. G. Chapman, T. M. Willey, A. M. Rubenchik, A. T. Anderson, Y. M. Wang, M. J. Matthews, W. E. King,  {\it Science} {\bf 368}, 660-665 (2020). 
\item[17.]
C. L. A. Leung, S. Marussi, R. C. Atwood, M. Towrie, P. J. Withers, P. D. Lee, {\it 24th Int. SFF Symp. - An Addit. Manuf. Conf.} {\bf Nat. Commun.}, 1–9 (2018). 
\item[18.]
R. Cunningham, C. Zhao, N. Parab, C. Kantzos, J. Pauza, K. Fezzaa, T. Sun, A. D. Rollett, {\it Science} {\bf 363}, 849–852 (2019). 
\item[19.]
S. J. Wolff, H. Wu, N. Parab, C. Zhao, K. F. Ehmann, T. Sun, J. Cao, {\it Sci. Rep.} {\bf 9}, 1–14 (2019).
\item[20.]
P. Raterron, S. Merkel, {\it J. Synchrotron Radiat. } {\bf 16}, 748–756 (2009).
\item[21.]
S. M. H. Hojjatzadeh, N. D. Parab, W. Yan, Q. Guo, L. Xiong, C. Zhao, M. Qu, L. I. Escano, X. Xiao, K. Fezzaa, W. Everhart, T. Sun, L. Chen, {\it Nat. Commun. } {\bf 10}, 1–8 (2019).
\item[22.]
M. Drakopoulos, T. Connolley, C. Reinhard, R. Atwood, O. Magdysyuk, N. Vo, M. Hart, L. Connor, B. Humphreys, G. Howell, S. Davies, T. Hill, G. Wilkin, U. Pedersen, A. Foster, N. De Maio, M. Basham, F. Yuan, K. Wanelik {\it J. Synchrotron Radiat.} {\bf 22}, 828–838 (2015).
\item[23.]
. Chen, S. J. Clark, C. Lun, A. Leung, L. Sinclair, S. Marussi, M. P. Olbinado, E. Boller, A. Rack, I. Todd, P. D. Lee, {\it Appl. Mater. Today.} {\bf 20}, 100650 (2020). 
\item[24.]
A. J. Pinkerton, L. Li, {\it Proc. Inst. Mech. Eng. Part B J. Eng. Manuf.} {\bf 218}, 363–374 (2004).
\item[25.]
J. C. Ion, H. R. Shercliff, M. F. Ashby, {\it Acta Metall. Mater.} {\bf 40}, 1539–1551 (1992).
\item[26.]
J. C. Haley, B. Zheng, U. S. Bertoli, A. D. Dupuy, J. M. Schoenung, E. J. Lavernia, {\it Mater. Des. } {\bf 161}, 86–94 (2019). 
\item[27.]
W. E. King, H. D. Barth, V. M. Castillo, G. F. Gallegos, J. W. Gibbs, D. E. Hahn, C. Kamath, A. M. Rubenchik,  {\it J. Mater. Process. Technol.} {\bf 214}, 2915–2925 (2014). 
\item[28.]
D. Dai, D. Gu, {\it Appl. Surf. Sci. } {\bf 355}, 310–319 (2015).
\item[29.]
F. Lia, J. Park, J. Tressler, R. Martukanitz, {\it Addit. Manuf.} {\bf 18}, 31–39 (2017).
\item[30.]
R. M. German, S. Activated, {\it Sintering: from Empirical Observations to Scientific Principles} (2014). 
\item[31.]
D. Li, H. He, J. Lou, Y. Li, Z. He, Y. Chen, F. Luo,  {\it Powder Technol.} (2019). 
\item[32.]
Z. A. Munir,{\it Powder Metall. } {\bf 24}, 177–180 (1981).
\item[33.]
R. Guillamet, J. Lopitaux, B. Hannoyer, M. Lenglet, {\it Le J. Phys. IV. } {\bf 03}, C9-349-C9-356 (1993).
\item[34.]
H. Taheri, M. R. B. M. Shoaib, L. W. Koester, T. A. Bigelow, P. C. Collins, L. J. Bond,  {\it Int. J. Addit. Subtractive Mater.} {\bf 1}, 172 (2017). 
\item[35.]
B. H. Rabin, G. R. Smolik, G. E. Korth,  {\it Metall. Mater. Trans. A.} {\bf 124}, 655-665 (2013). 
\item[36.]
X. Lin, T. M. Yue, H. O. Yang, W. D. Huang,  {\it Metall. Mater. Trans. A.} {\bf 38}, 127–137 (2007).
\item[37.]
S. Gorsse, C. Hutchinson, M. Gouné, R. Banerjee,  {\it Sci. Technol. Adv. Mater.} {\bf 18}, 584–610 (2017). 
\item[38.]
C. Beckermann, K. D. Carlson,{\it Sensitivity of Steel Casting Simulation Results to Alloy Property Dataset}, 1–27 (2012). 
\end{itemize}

\clearpage

\section*{Supplementary materials}
\subsection*{1. Supplementary materials and methods}
\textbf{Synchrotron Imaging Conditions:} We performed \emph{in situ} and \emph{operando} X-ray imaging on the $I12-JEEP$ beamline at Diamond Light Source to capture the transient phenomena of the DED-AM of SS316 and Ti6242. A monochromatic beam was selected to ensure linear attenuation. A mean X-ray energy of approximately 53 $keV$ was used for all experiments. The X-ray imaging system consisted of a 200 $\mu m$ thick LuAg:Ce scintillator and a 4× magnification long working distance objective lens (0.21 numerical aperture). The X-ray images were captured by a high-resolution imaging camera (PCO.edge 5.5, PCO) at 200 and at up to 2,000 $fps$ using a lower resolution high-speed camera (MIRO 310M, Vision Research Inc.). Unlike laser powder bed fusion, the weld pool in DED-AM is much larger and the translations are much slower ($mm\ s^{-1}$ compared to $m\ s^{-1}$). Therefore, we optimised the number of frames per second for two situations. Firstly, to capture pool phenomena we found 200 $fps$ was sufficient to capture the key phenomena, while enabling a long exposure time of 0.0049 $s$ and a very small off-load time (0.0001 $s$) that gave excellent signal to noise ($S/N$) ratio. We also used 2000 $fps$ for experiments where faster phenomena were occurring, trading off $S/N$ for speed.\\

\subsection*{2. Supplementary figures}

\setcounter{figure}{0}
\makeatletter 
\renewcommand{\thefigure}{S\@arabic\c@figure}
\makeatother

\begin{figure}[H]
\includegraphics[width=15cm]{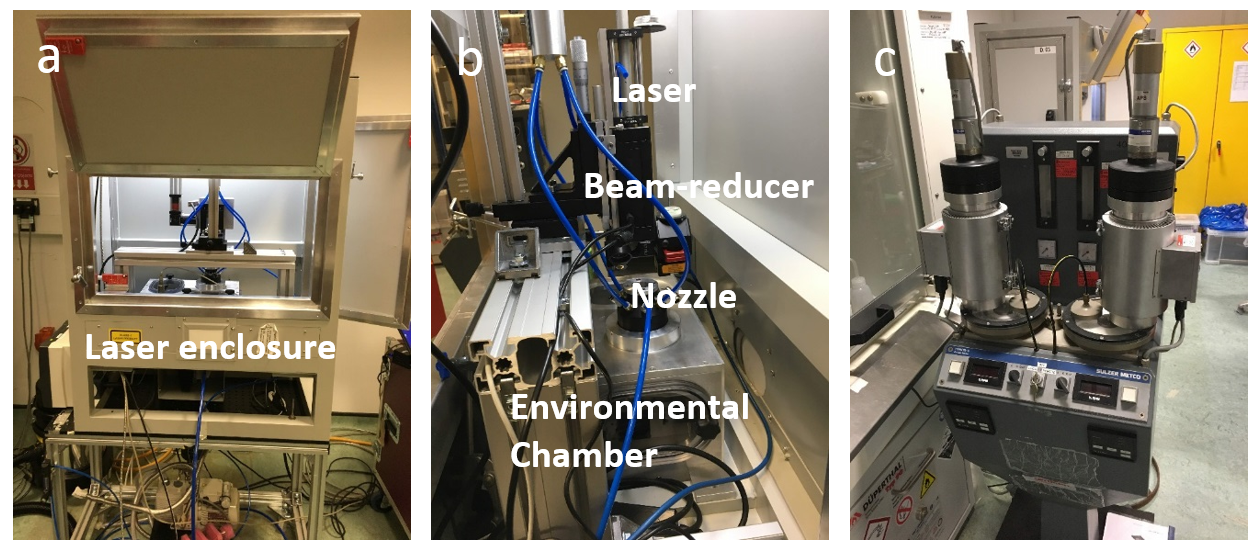}
\centering
\caption{ \textbf{Experimental setup for the in situ laser additive manufacturing process using Blown Powder Additive Manufacturing Process Replicator (BAMPR). } (A) Class I laser enclosure. (B) DED process replicator mounted inside the laser enclosure. (C) Powder feeder. }
\end{figure}

\begin{figure}[H]
\includegraphics[width=15cm]{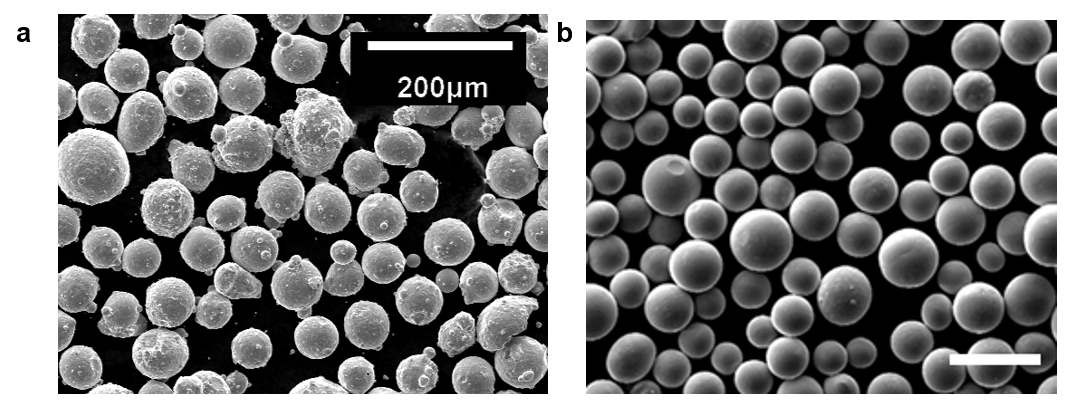}
\centering
\caption{ \textbf{Powder analysis. } SEM image of (A) SS316L powder. Scale bar = 200 $\mu m$; (B) Ti6242 powder. Scale bar = 100$\mu m$. Gas atomised powders was used in the experiments with a size distribution of 45 - 90 $\mu m$ and a $D_{50}$ (median diameter) of 70 $\mu m$. Particle size distribution is analysed using same method in Leung et al’s paper({\it 1}). }
\end{figure}

\begin{figure}[H]
\includegraphics[width=15cm]{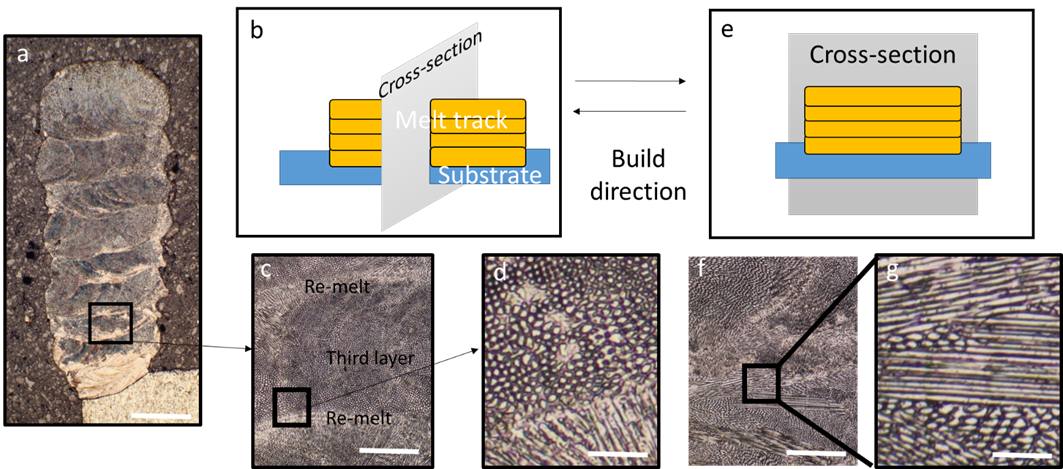}
\centering
\caption{ \textbf{Microstructure analysis of DED-AM SS316 samples from the imaging experiment. } Optical microscope image of SS316L DED-AM microstructure from BAMPR experiment under $P$ = 6.37 × 10$^{3}$ $W\ mm^{-2}$, scanning speed $v$ = 1.67 $mm\ s^{-1}$. Scale bar = 250 $\mu m$; (A) cross-section of a ten layer thin wall build and the schematic of the cross-section direction is shown in (B). Scale bar = 1 $mm$. (C) Enlarged view of (A) showing microstructure of the third layer and the surrounding re-melt layer. Scale bar = 500 $\mu m$.  (D) Enlarged view of (C) showing detailed microstructure of the melt track and re-melt area. Grain size is $\sim$ 10 $\mu m$. Scale bar = 50 $\mu m$. (F) Cross-section along the build direction of the same sample and the schematic of the cross-section direction is shown in (E). Scale bar = 500 $\mu m$. (G) Enlarged view of (F) showing detailed microstructure of the melt track. Scale bar = 50 $\mu m$. Elongated grains following the laser path can be observed. }
\end{figure}

\begin{figure}[H]
\includegraphics[width=15cm]{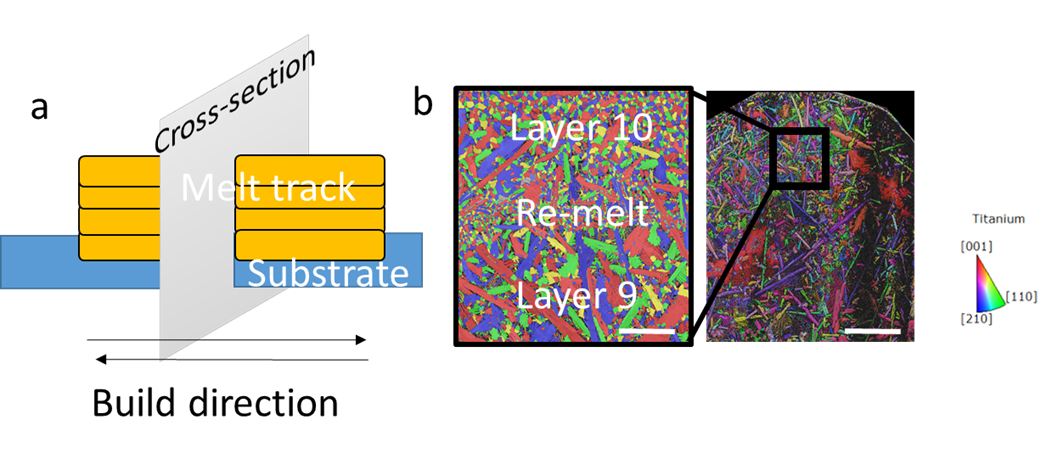}
\centering
\caption{ \textbf{Microstructure analysis of DED-AM Ti6242 samples from the imaging experiments. } EBSD images of SS316L DED-AM microstructure from BAMPR experiment under $P$ = 1.59 × 10$^{3}$ $W\ mm^{-2}$, $v$ = 1 $mm\ s^{-1}$. Cross-section along the build direction. Long columnar grains and small equiaxed grain in the re-melt area can be observed. Scale bar = 250 $\mu m$ and 100 $\mu m$ for normal view and expanded view, respectively. }
\end{figure}

\begin{figure}[H]
\includegraphics[width=14cm]{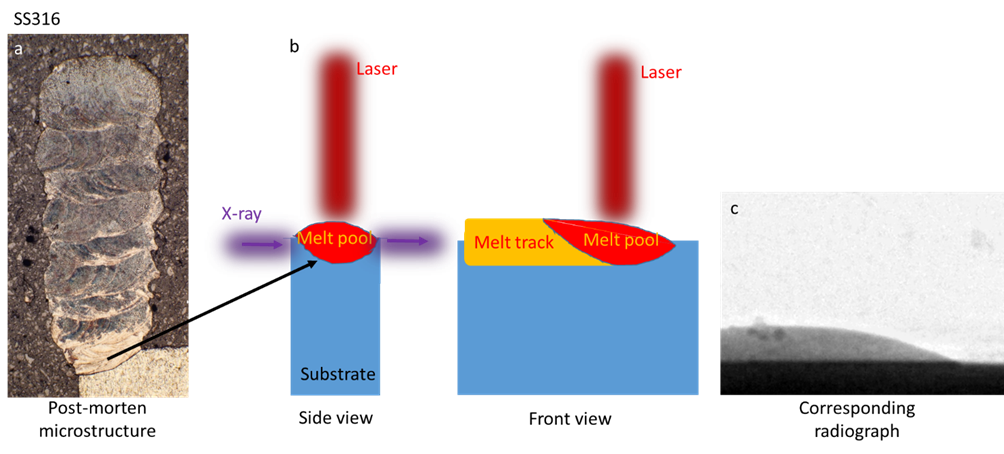}
\centering
\includegraphics[width=14cm]{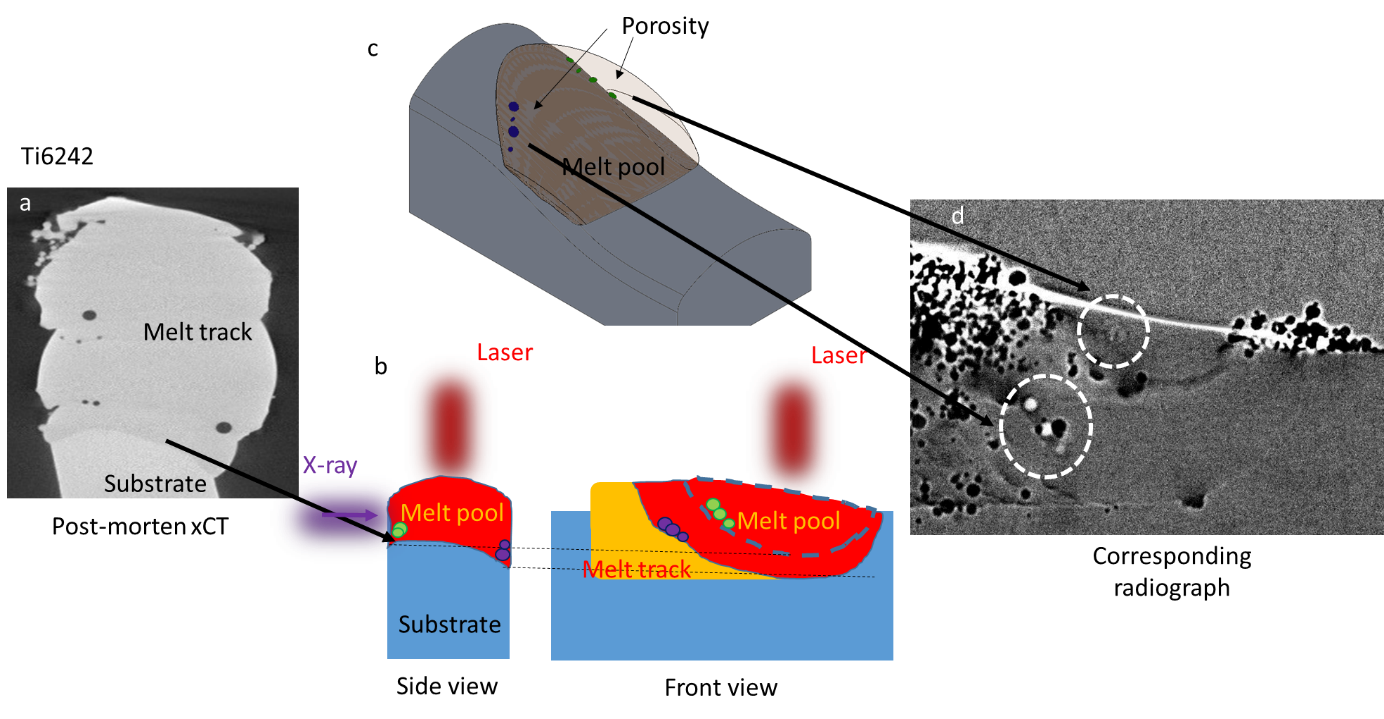}
\centering
\caption{ \textbf{Melt pool morphology of DED-AM from the imaging experiments.  }The melt pool area is defined as liquid and solid-liquid boundary is marked by observing the moving boundaries. A double melt pool sometimes can be observed showing moving porosity outside of the defined melt pool area during Ti6242 build, as shown in the radiograph revealed by local-temporal background subtraction. It is due to a concave shape at the bottom of the melt pool, as shown in the cross-section view of the post-mortem xCT scan. In Ti6242: (A) Post-mortem xCT cross-section shows the concave shape of the bottom of the melt pool. (B) Schematic shows the mechanism behind double melt pool under X-ray imaging. (C) 3D schematic shows the porosity in the double melt pool are sitting on the two edges of the concave melt pool edge. (D) Corresponding filtered radiograph shows the boundary of double melt pool and the porosity inside.  In the contrary, the convex shape of the bottom of the melt pool in SS316 results in a single tear drop shape melt pool: (A) Post-mortem microstructure analysis reveal the convex shape of the bottom of the melt pool. (B) Schematic shows the mechanism behind single tear drop shape melt pool. (C) Corresponding radiograph. }
\end{figure}

\begin{figure}[H]
\includegraphics[width=8cm]{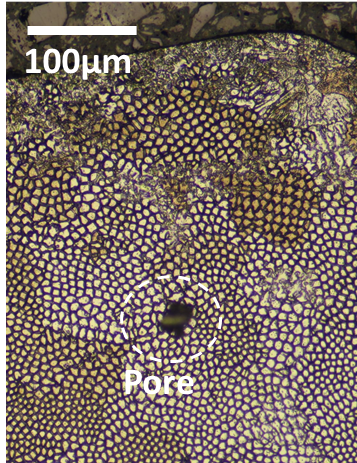}
\centering
\caption{ \textbf{Porosity analysis in SS316.} Optical microscope image of a cross-section of SS316 build. It reveals a gas entrapped pore embedded in the cellular colonies. $P$ = 6.37 × 10$^{3}$ $W\ mm^{-2}$, $v$ = 1.67 $mm\ s^{-1}$. }
\end{figure}

\begin{figure}[H]
\includegraphics[width=10cm]{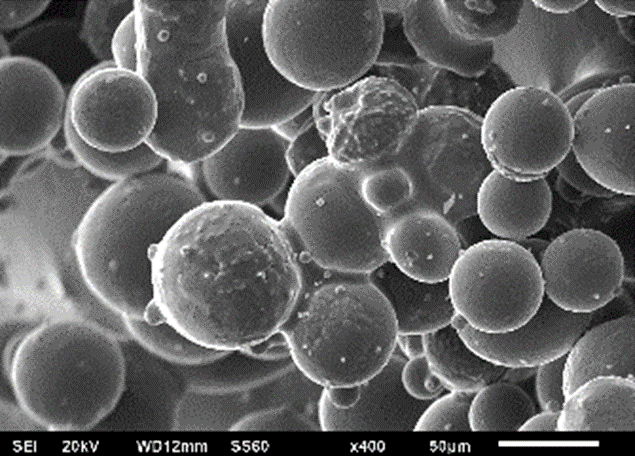}
\centering
\caption{ \textbf{Power agglomeration in Ti6242.} Optical microscope image of a cross-section of SS316 build. It reveals a gas entrapped pore embedded in the cellular colonies. $P$ = 1.59 × 10$^{3}$ $W\ mm^{-2}$, $v$ = 1 $mm\ s^{-1}$. }
\end{figure}

\begin{figure}[H]
\includegraphics[width=14cm]{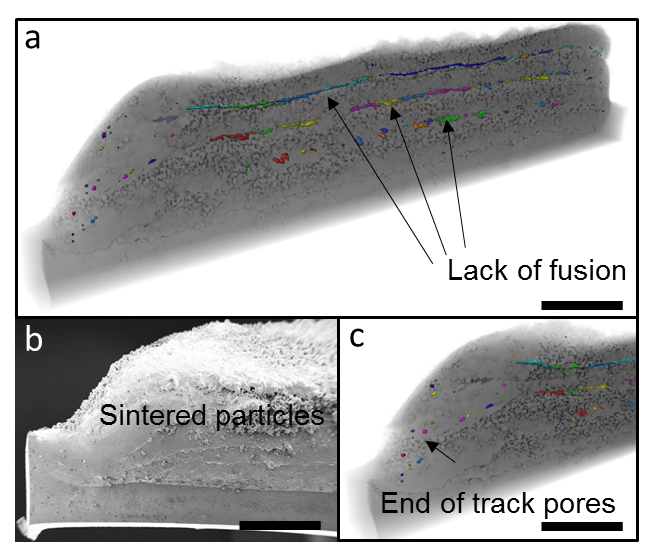}
\centering
\caption{ \textbf{Porosity and lack of fusion features in Ti6242.} Micro-CT and SEM scans reveal (a) lack of fusion between layers, (b) sintered particles on top of the track, and (b) end of track pores. $P$ = 1.59 × 10$^{3}$ $W\ mm^{-2}$, $v$ = 1 $mm\ s^{-1}$. Scale bar = 1 $mm$.}
\end{figure}

\subsection*{3. Supplementary tables}

\makeatletter 
\renewcommand{\thetable}{S\@arabic\c@table}
\makeatother

\begin{table}[H]
\centering
\begin{tabular}{|p{1.4cm}|p{1.4cm}|p{1.4cm}|p{1.5cm}|p{1.5cm}|p{1.5cm}|p{1.5cm}|p{1.4cm}|p{1.4cm}|} \hline
 \multicolumn{9}{|c|}{\textbf{Powder material: SS316L}} \\ \hline
 Test No. & Spot size ($\mu m$)&Power ($W$)&Scan speed ($mm\ s^{-1}$)&Shield gas flowrate ($l\ min^{-1}$) &Carrier gas flowrate ($l\ min^{-1}$)&Powder feedrate ($g\ min^{-1}$)&Hatch&Camera \\  \hline
SS1  & 200 & 200 & 1.67 & 6 & 6 & 5 & 1 & Miro  \\  \hline
SS2  & 200 & 200 & 1.67 & 6 & 6 & 5 & 1 & Miro  \\  \hline
SS3  & 200 & 200 & 1.67 & 6 & 6 & 5 & 1 & Miro  \\  \hline
SS4  & 200 & 200 & 1.67 & 6 & 6 & 5 & 1 & Miro  \\  \hline
SS5  & 200 & 200 & 10 & 6 & 6 & 5 & 1 & Miro  \\  \hline
SS6  & 200 & 200 & 5 & 6 & 6 & 5 & 1 & Miro  \\  \hline
SS7  & 200 & 200 & 3 & 6 & 6 & 5 & 1 & Miro  \\  \hline
SS8  & 200 & 200 & 1.67 & 8 & 6 & 5 & 1 & Miro  \\  \hline
SS9  & 200 & 200 & 1.67 & 0 & 8 & 5 & 1 & Miro  \\  \hline
SS10  & 200 & 200 & 1.67 & 6 & 2 & 5 & 1 & Miro  \\  \hline
SS11  & 200 & 200 & 1.67 & 6 & 6 & 3 & 1 & Miro  \\  \hline
SS12  & 200 & 200 & 1.67 & 6 & 6 & 3.5 & 1 & Miro  \\  \hline
SS13  & 200 & 200 & 1.67 & 6 & 6 & 6 & 1 & Miro  \\  \hline
SS14  & 200 & 200 & 1.67 & 6 & 6 & 7.5 & 1 & Miro  \\  \hline
\end{tabular}
\caption{\textbf{Experimental list using SS316L } Listed parameters including laser power, laser spot size, gas flow rate, powder feedrate and substrate traverse speed.}
\label{table:1}
\end{table}

\begin{center}
\begin{longtable}{|p{1.4cm}|p{1.4cm}|p{1.4cm}|p{1.5cm}|p{1.5cm}|p{1.5cm}|p{1.5cm}|p{1.4cm}|p{1.4cm}|} \hline
\multicolumn{9}{|c|}{\textbf{Powder material: Ti6242}} \\ \hline
\endfirsthead
\multicolumn{9}{c}{{\bfseries \tablename\ \thetable -- -- continued from previous page}} \\
\endhead
\hline \multicolumn{9}{|r|}{{Continued on next page}} \\ \hline
\endfoot
\endlastfoot
Test No. & Spot size ($\mu m$)&Power ($W$)&Scan speed ($mm\ s^{-1}$)&Shield gas flowrate ($l\ min^{-1}$) &Carrier gas flowrate ($l\ min^{-1}$)&Powder feedrate ($g\ min^{-1}$)&Hatch&Camera \\  \hline
Ti1  & 200 & 200 & 1 & 6 & 6 & 0.25 & 1 & Edge  \\  \hline
Ti2  & 200 & 200 & 1 & 6 & 6 & 1 & 1 & Edge  \\  \hline
Ti3  & 400 & 200 & 0.5 & 6 & 6 & 0.125 & 1 & Miro  \\  \hline
Ti4  & 200 & 200 & 5 & 6 & 6 & 0.125 & 1 & Miro  \\  \hline
Ti5  & 200 & 200 & 5 & 6 & 6 & 0.125 & 1 & Miro  \\  \hline
Ti6  & 200 & 200 & 5 & 6 & 6 & 0.125 & 1 & Miro  \\  \hline
Ti7  & 200 & 200 & 5 & 6 & 6 & 0.125 & 1 & Miro  \\  \hline
Ti8  & 200 & 200 & 5 & 6 & 6 & 0.125 & 1 & Miro  \\  \hline
Ti9  & 200 & 200 & 5 & 6 & 6 & 0.125 & 1 & Miro  \\  \hline
Ti10  & 200 & 200 & 5 & 6 & 6 & 0.125 & 1 & Miro  \\  \hline
Ti11  & 200 & 200 & 5 & 6 & 6 & 0.125 & 1 & Miro  \\  \hline
Ti12  & 200 & 200 & 5 & 6 & 6 & 0.125 & 1 & Miro  \\  \hline
Ti13  & 200 & 200 & 5 & 6 & 6 & 0.125 & 1 & Miro  \\  \hline
Ti14  & 200 & 200 & 5 & 6 & 6 & 0.125 & 1 & Miro  \\  \hline
Ti15  & 200 & 200 & 5 & 6 & 6 & 0.125 & 1 & Miro  \\  \hline
Ti16  & 200 & 200 & 5 & 6 & 6 & 0.125 & 1 & Miro  \\  \hline
Ti17  & 200 & 200 & 5 & 6 & 6 & 0.125 & 1 & Miro  \\  \hline
Ti18  & 200 & 200 & 5 & 6 & 6 & 0.125 & 1 & Miro  \\  \hline
Ti19  & 200 & 200 & 5 & 6 & 6 & 0.125 & 1 & Miro  \\  \hline
Ti20  & 200 & 200 & 5 & 6 & 6 & 0.125 & 1 & Miro  \\  \hline
Ti21  & 200 & 200 & 5 & 6 & 6 & 0.125 & 1 & Miro  \\  \hline
Ti22  & 200 & 200 & 5 & 6 & 6 & 0.125 & 1 & Miro  \\  \hline
Ti23  & 200 & 200 & 5 & 6 & 6 & 0.125 & 1 & Miro  \\  \hline
Ti24  & 200 & 200 & 5 & 6 & 6 & 0.125 & 1 & Miro  \\  \hline
Ti25  & 200 & 200 & 5 & 6 & 6 & 0.125 & 1 & Miro  \\  \hline
Ti26  & 200 & 200 & 5 & 6 & 6 & 0.125 & 1 & Miro  \\  \hline
Ti27  & 200 & 200 & 5 & 6 & 6 & 0.125 & 1 & Miro  \\  \hline
Ti28  & 200 & 200 & 5 & 6 & 6 & 0.125 & 1 & Miro  \\  \hline
Ti29  & 200 & 200 & 5 & 6 & 6 & 0.125 & 1 & Miro  \\  \hline
Ti30  & 200 & 200 & 5 & 6 & 6 & 0.125 & 1 & Miro  \\  \hline
Ti31  & 200 & 200 & 5 & 6 & 6 & 0.125 & 1 & Miro  \\  \hline
Ti32  & 200 & 200 & 5 & 6 & 6 & 0.125 & 1 & Miro  \\  \hline
Ti33  & 200 & 200 & 5 & 6 & 6 & 0.125 & 1 & Miro  \\  \hline
Ti34  & 200 & 200 & 5 & 6 & 6 & 0.125 & 1 & Miro  \\  \hline
Ti35  & 200 & 200 & 5 & 6 & 6 & 0.125 & 1 & Miro  \\  \hline
Ti36  & 200 & 200 & 5 & 6 & 6 & 0.125 & 1 & Miro  \\  \hline
Ti37  & 200 & 200 & 5 & 6 & 6 & 0.125 & 1 & Miro  \\  \hline
Ti38  & 200 & 200 & 5 & 6 & 6 & 0.125 & 1 & Miro  \\  \hline
Ti39  & 200 & 200 & 5 & 6 & 6 & 0.125 & 1 & Miro  \\  \hline
Ti40  & 200 & 200 & 5 & 6 & 6 & 0.125 & 1 & Miro  \\  \hline
\end{longtable}
\end{center}
Table S2:\textbf{Experimental list using Ti6242.} Listed parameters including laser power, laser spot size, gas flow rate, powder feedrate and substrate traverse speed.

\begin{table}[H]
\centering
\begin{tabular}{|p{1.8cm}|p{1.4cm}|p{2.2cm}|p{2.2cm}|p{2cm}|p{1.8cm}|p{2.0 cm}|} \hline
\textbf{Laser power} & \textbf{Spot size} &\textbf{Power density} &\textbf{Scan speed} & \textbf{Powder feedrate} & \textbf{Energy density} & \multirow{2}{*}{\textbf{References}} \\  \cline{1-6}
($W$)  & ($mm$) & ($W\ mm^{-2}$) & ($mm\ s^{-1}$) & ($g\ s^{-1}$) & ($J\ mm^{-3}$)&   \\  \hline
1400  & 3 & 198 & 16.6 & 0.23 & 19.40 & ({\it 2})  \\  \hline
130-210  & 0.1 & 1655-2674 & 300-1300 &  & 33-233  & ({\it 3})  \\  \hline
2300,4000  &  &  & 8.5,23  &  0.3, 0.38   &81, 66& ({\it 4})  \\  \hline
100-3000  &  &  & 5-20 & 0.1-1  & 2-150 & ({\it 5})  \\  \hline
50-200  & 0.2, 0.4 & 398-6366 & 1-5 & 0.017-0.16 & 0.16-32  & Our work  \\  \hline
\end{tabular}
\caption{\textbf{Laser energy density and power density. } Experimental parameters in comparison to literature.}
\label{table:3}
\end{table}

\begin{table}[H]
\centering
\begin{tabular}{|p{1.8cm}|p{1.8cm}|p{1.3cm}|p{1.8cm}|p{2.1cm}|p{2.1cm}|p{2.1 cm}|} \hline
\multirow{2}{*}{\textbf{Materials}} & \textbf{Laser power} &\textbf{Scan speed} &\textbf{Powder feedrate} & \textbf{Melt pool volume} & \textbf{Melt pool length} & \textbf{Melt pool depth} \\  \cline{2-7}
 &  ($mm\ s^{-1}$)   & ($g\ s^{-1}$)  & ($mm^{3}$) & ($\mu m$) & ($\mu m$)  & ($\mu m$) \\ \hline
\multirow{7}{*}{\textbf{SS316L}} & 100 & 1.67 & 5 & 0.65 $\pm$ 0.08 & 1759 $\pm$ 13 & 514 $\pm$ 5  \\  \cline{2-7}
 & 150 & 1.67 & 5 & 1.41 $\pm$ 0.17  & 2215 $\pm$ 17 & 713 $\pm$ 4  \\  \cline{2-7}
 & 200 & 1.67 & 5 & 1.64 $\pm$ 0.11 & 2356 $\pm$ 11 & 719 $\pm$ 6  \\  \cline{2-7}
 & 200 & 3.3 & 5 & 0.61 $\pm$ 0.02 & 1836 $\pm$ 9 & 447 $\pm$ 3  \\  \cline{2-7}
 & 200 & 5 & 5 & 0.49 $\pm$ 0.02 & 1832 $\pm$ 8 & 514 $\pm$ 3  \\  \cline{2-7}
 & 200 & 1.67 & 3 & 0.85 $\pm$ 0.01 & 2168 $\pm$ 3 & 431 $\pm$ 2  \\  \cline{2-7}
 & 200 & 1.67 & 7.5 & 4.8 $\pm$ 0.12 & 3000 $\pm$ 41 & 1187 $\pm$ 21  \\  \hline
\multirow{6}{*}{\textbf{Ti6242}} & 100 & 1 & 1 & N/A & 148 $\pm$ 11 & 27 $\pm$ 3  \\  \cline{2-7}
 & 200 & 1 & 1 & N/A & 500 $\pm$ 21 & 105 $\pm$ 7  \\  \cline{2-7}
 & 200 & 2.5 & 1 & N/A & 493 $\pm$ 18 & 178 $\pm$ 5  \\  \cline{2-7}
 & 200 & 5 & 1 & N/A & 351 $\pm$ 9 & 18 $\pm$ 1  \\  \cline{2-7}
 & 200 & 1 & 0.5 & N/A & 576 $\pm$ 10 & 189 $\pm$ 1  \\  \cline{2-7}
 & 200 & 1 & 2 & N/A & 306 $\pm$ 32 & 98 $\pm$ 6  \\  \hline
\end{tabular}
\caption{\textbf{Melt pool geometry measured from experiments. } Melt pool volumes and profiles variation across powder feed-rate, laser power and traverse speed. Error is $SD$. SS316 melt pool is regarded as an elliptical-cap shape for the volume calculation. The melt pool length and depth of Ti6242 melt pool is measured as the bigger value from the double melt pool shape. Due to the complex melt pool shape of the TI6242, its volume is not calculated. }
\label{table:4}
\end{table}

\subsection*{4. Supplementary movies}
 \noindent Movie S1 \textbf{Time-resolved radiographs acquired during DED-AM at the start of a melt track on SS316. } SS316 build is under laser power density $P$ = 6.37 × 10$^{3}$ $W\ mm^{-2}$, scanning speed $v$ = 1.67 $mm\ s^{-1}$, powder feedrate 3 $g\ min^{-1}$, captured at 200 $fps$. Experiment Number $SS11$. \\ \\
 \noindent Movie S2 \textbf{Time-resolved radiographs acquired during DED-AM at the start of a melt track on Ti6242. } Ti6242 build is under $P$ = 1.59 × 10$^{3}$ $W\ mm^{-2}$, $v$ = 1 $mm\ s^{-1}$, powder feedrate 1 $g\ min^{-1}$, captured at 200 $fps$. Experiment Number $Ti2$. \\ \\
 \noindent Movie S3 \textbf{Time-series radiographs acquired during DED of a SS316 multi-layer build. } SS316 build is under laser power density $P$ = 6.37 × 10$^{3}$ $W\ mm^{-2}$, scanning speed $v$ = 1.67 $mm\ s^{-1}$, powder feedrate 3 $g\ min^{-1}$, captured at 200 $fps$. Experiment Number $SS11$. \\ \\
 \noindent Movie S4 \textbf{Time-series radiographs acquired during DED of a Ti6242 multi-layer build. } Ti6242 build is under $P$ = 1.59 × 10$^{3}$ $W\ mm^{-2}$, $v$ = 1 $mm\ s^{-1}$, powder feedrate 1 $g\ min^{-1}$, captured at 200 $fps$. Experiment Number $Ti2$.  \\ \\
 \noindent Movie S5 \textbf{The initial establishment of a melt track at the onset of a build was captured to reveal the mechanism at laser powers of 50 $W$, $P$ = 1.59 × 10$^{3}$ $W\ mm^{-2}$. } Scanning speed $v$ = 1.67 $mm\ s^{-1}$, powder feedrate 3 $g\ min^{-1}$, captured at 200 $fps$. Experiment Number $SS4$. \\ \\
 \noindent Movie S6 \textbf{The initial establishment of a melt track at the onset of a build was captured to reveal the mechanism at laser powers of 100 $W$, $P$ = 3.18 × 10$^{3}$ $W\ mm^{-2}$. } Scanning speed $v$ = 1.67 $mm\ s^{-1}$, powder feedrate 3 $g\ min^{-1}$, captured at 200 $fps$. Experiment Number $SS3$. \\ \\
 \noindent Movie S7 \textbf{The initial establishment of a melt track at the onset of a build was captured to reveal the mechanism at laser powers of 200 $W$, $P$ = 6.37 × 10$^{3}$ $W\ mm^{-2}$. } Scanning speed $v$ = 1.67 $mm\ s^{-1}$, powder feedrate 3 $g\ min^{-1}$, captured at 200 $fps$. Experiment Number $SS1$. \\ \\
 \noindent Movie S8 \textbf{The sintering phenomenon during DED-AM of Ti6242.  } The sintered layer thickness is more than three times the height of the melt track itself. $P$ = 7.96 × 10$^{2}$ $W\ mm^{-2}$, $v$ = 1 $mm\ s^{-1}$, powder feedrate 1 $g\ min^{-1}$, captured at 200 $fps$. Experiment Number $Ti18$. \\ \\
 \noindent Movie S9 \textbf{SS316: A pore appears to form via gas entrainment from agglomerated powder particles. } A pore flow trajectory revealed on radiographs. SS316 build is under laser power density $P$ = 1.59 × 10$^{3}$ $W\ mm^{-2}$, scanning speed $v$ = 1.67 $mm\ s^{-1}$, powder feedrate 3 $g\ min^{-1}$, captured at 200 $fps$. Experiment Number $SS11$.\\ \\
 \noindent Movie S10 \textbf{Ti6242: pores are shown being pushed along in the melt pool. } SS316 build is under laser power density $P$ = 6.37 × 10$^{3}$ $W\ mm^{-2}$, scanning speed $v$ = 1.67 $mm\ s^{-1}$, powder feedrate 3 $g\ min^{-1}$, captured at 200 $fps$. Experiment Number $SS11$. \\ \\

\subsection*{References}

\begin{enumerate}
\setcounter{enumi}{0}
\item[1.]
C. L. A. Leung, S. Marussi, R. C. Atwood, M. Towrie, P. J. Withers, P. D. Lee, {\it Nat. Commun.} {\bf 9}, 1-9 (2019). 
\item[2.]
Z. E. Tan, J. H. L. Pang, J. Kaminski, H. Pepin,  {\it Addit. Manuf. } {\bf 25}, 286-296 (2019). 
\item[3.]
Z. A. Mierzejewska, {\it Materials.} {\bf 12}, 2331 (2019). 
\item[4.]
Z. Wang, T. A. Palmer, A. M. Beese,  {\it Acta Mater.} {\bf 110}, 226–235 (2016). 
\item[5.]
T. Debroy, H.L. Wei, J.S. Zuback, T. Mukherjee, J.W. Elmer, J.O. Milewski, A.M. Beese, A. Wilson-Heid, A. De, W. Zhang, {\it Prog. Mater. Sci.} {\bf 92}, 112-224 (2018). 
\end{enumerate}

\end{document}